\documentclass[runningheads]{llncs}
\usepackage{dialogue}
\usepackage{tabularx}
\usepackage{amssymb}
\usepackage{amsmath} 
\usepackage{algorithm}
\usepackage{algpseudocode}
\usepackage{float}
\usepackage{mdframed}
\usepackage{xcolor}
\usepackage{threeparttable}
\usepackage[T1]{fontenc}
\usepackage{graphicx}
\usepackage{enumitem}
\usepackage{booktabs}
\newcommand{\revision}[1]{\textcolor{black}{#1}}
\newcommand{\revisiontwo}[1]{\textcolor{black}{#1}}

\usepackage[protrusion]{microtype} 
\usepackage[breaklinks,colorlinks,linkcolor=blue,urlcolor=blue,citecolor=blue]{hyperref} 

\newcommand{\numteamstotal}{24}        
\newcommand{\numteamsanalyzed}{23}     
\newcommand{\numstudentstotal}{81}     
\newcommand{\numstudentsanalyzed}{76}  

\usepackage{tikz}
\newcommand\copyrighttext{%
  \scriptsize This version of the article has been accepted for publication, after peer review, but is not the Version of Record and does not reflect post-acceptance improvements, or any corrections. The Version of Record is published in the \textit{27th International Conference on Artificial Intelligence in Education (AIED 2026)}, and is available online at \url{https://doi.org/10.1007/978-3-032-29760-0_47}.
  \smallskip
  
  \textcopyright\ 2026. Please cite this article as follows: C. Borchers, V. Švábenský, S. Kafle, K. Tang, J. Vykopal: \textit{Multimodal Analytics of Cybersecurity Crisis Preparation Exercises: What Predicts Success?}. In 27th International Conference on Artificial Intelligence in Education (AIED 2026), Springer Nature, 2026. DOI: \href{https://doi.org/10.1007/978-3-032-29760-0_47}{10.1007/978-3-032-29760-0\_47}.}
\newcommand\copyrightnotice{%
\begin{tikzpicture}[remember picture,overlay]
\node[anchor=north,yshift=-24pt] at (current page.north) {\fbox{\parbox{\dimexpr\textwidth-\fboxsep-\fboxrule\relax}{\copyrighttext}}};
\end{tikzpicture}%
}

\begin{document}
\title{Multimodal Analytics of Cybersecurity Crisis Preparation Exercises: What Predicts Success?}
\titlerunning{Multimodal Analytics of Cybersecurity Crisis Preparation Exercises}

\author{Conrad Borchers\inst{1}\orcidID{0000-0003-3437-8979} \and \\
Valdemar Švábenský\inst{2}\orcidID{0000-0001-8546-280X} \and \\
Sandesh K. Kafle\inst{2}\orcidID{0009-0002-4275-1171} \and \\
Kevin K. Tang\inst{1}\orcidID{0009-0009-4544-7467} \and \\
Jan Vykopal\inst{2}\orcidID{0000-0002-3425-0951}}
\authorrunning{C. Borchers et al.}

\institute{Carnegie Mellon University\\
\email{cborcher@cs.cmu.edu, kktang@andrew.cmu.edu}\\
\and
Masaryk University, Faculty of Informatics\\
\email{valdemar@mail.muni.cz, \{xkafle,vykopal\}@fi.muni.cz}}

\maketitle 
\begin{abstract}
Instructional alignment, the match between intended cognition and enacted activity, is central to effective instruction but hard to operationalize at scale. We examine alignment in cybersecurity simulations using multimodal traces from 23 teams (76 students) across five exercise sessions. Study 1 codes objectives and team emails with Bloom's taxonomy and models the completion of key exercise tasks with generalized linear mixed models. Alignment, defined as the discrepancy between required and enacted Bloom levels, predicts success, whereas the Bloom category alone does not predict success once discrepancy is considered. Study 2 compares predictive feature families using grouped cross-validation and $\ell_{1}$-regularized logistic regression. Text embeddings and log features outperform Bloom-only models (AUC~$\approx$~0.74 and 0.71 vs. 0.55), and their combination performs best (Test AUC~$\approx$~0.80), with Bloom frequencies adding little. Overall, the work offers a measure of alignment for simulations and shows that multimodal traces best forecast performance, while alignment provides interpretable diagnostic insight.

\copyrightnotice
\end{abstract}

\keywords{Bloom, simulation-based learning, tabletop exercises, prediction}

\section{Introduction}

Instructional alignment, the coherence among learning objectives, activities, and assessment, is foundational to effective learning design \cite{biggs1996enhancing,porter2002measuring}. Bloom's taxonomy and later revisions \cite{bloom1956taxonomy,10.1145/3587276} provide a shared language for articulating cognitive demands, ranging from remembering and understanding to analyzing, evaluating, and creating. Despite their ubiquity in instructional design, such frameworks are rarely used to \emph{measure} alignment in practice, particularly at scale, where learners' enacted activities may diverge from intended objectives~\cite{tello2020novel}. \revision{The present study contributes empirical evidence that alignment between collaborative learner actions and instructional objectives, inferred from multimodal learning analytics, predicts problem-solving performance. For the AIED community, this demonstrates how theory-driven cognitive frameworks can be operationalized to support performance prediction and complex collaborative learning environments.}

This challenge is especially pronounced in simulation-based learning (SBL), where open-ended, team-based tasks generate rich but noisy multimodal traces that are difficult to relate to targeted cognitive processes \cite{chernikova2020simulation,Yan2023simulation}. Cybersecurity tabletop exercises (TTXs; see Section~\ref{subsec:related-work-TTX}) exemplify this problem: teams must coordinate under time pressure, interpret evolving information, and communicate with diverse stakeholders, producing behaviors that are pedagogically valuable yet methodologically hard to code and assess \cite{Vykopal2024research,Svabensky2024from}. Although past work has modeled such traces to support instructional intervention \cite{wise2014designing}, it remains unclear whether Bloom-aligned discrepancies between objectives and enacted behavior predict performance, or which trace types are most informative.

We address these questions using cybersecurity TTXs conducted on the open-source INJECT Exercise Platform \cite{Svabensky2024from}, analyzing platform logs and in-exercise email communications from \numteamsanalyzed\ teams (\numstudentsanalyzed\ students) across five sessions. RQ1 examines whether \emph{alignment}, operationalized as the Bloom-level discrepancy between milestone objectives (i.e., the successful completion of key tasks in the exercise) and the highest Bloom level evidenced in team communications, predicts \revision{team performance}, while RQ2 compares the predictive utility of Bloom-coded categories, linguistic features from emails and logs, and engineered behavioral log features, individually and in combination. Study~1 codes objectives and communications using Bloom's taxonomy and applies generalized linear mixed models~\cite{bates2015fitting} to test the contribution of discrepancy beyond raw Bloom levels, and Study~2 evaluates predictive models using Bloom-only features, text embeddings, behavioral aggregates, and their combinations with grouped cross-validation and $\ell_{1}$-regularized logistic regression.

We contribute a data-driven measure of instructional alignment in SBL and show that Bloom-level discrepancy between objectives and team communication predicts performance in cybersecurity tabletop exercises. Linguistic and behavioral traces were more predictive than Bloom-frequency features, while discrepancy remains valuable for instructional diagnosis.

\section{Background}

\subsection{Bloom's Taxonomy for the Prediction of Learning Performance}

Bloom's Taxonomy is an established framework for characterizing learning objectives across cognitive levels, including in computing education \cite{bloom1956taxonomy,10.1145/3587276}. AIED research has operationalized this framework by inferring cognitive levels from observable learner behaviors, such as clickstream activity, assignment artifacts, and rubric-based assessments, to predict performance and cognitive progression~\cite{Sori_Mustapha_2025,li2022automatic,Ayyanathan_2022}. Despite their success, these approaches rely primarily on unimodal digital traces, which limits their ability to capture learning processes distributed across multiple forms of interaction.

Recent AI-oriented reinterpretations, including the proposed ``AIEd Bloom's taxonomy,'' replace cognitive levels with process-focused stages such as Collect and Adapt \cite{Mohammad_Shaqour}. While motivated by modernization, this shift weakens Bloom's theoretical grounding by obscuring distinctions between lower- and higher-order thinking and offering limited support for differentiating surface engagement from critical reasoning. Both traditional and AI-driven adaptations largely overlook the application of Bloom's framework to multimodal learning contexts, including collaborative communication artifacts, which require explicit coding schemes and ground-truth labeling \cite{li2022automatic}. We address these gaps by combining multimodal evidence from team communication transcripts and interaction logs with predictive modeling to operationalize Bloom's Taxonomy in a rich learning setting.
\revisiontwo{To address validity concerns in mapping trace data to cognition, we interpret email and log features as indicators of collaborative processes such as coordination and shared problem framing, which have been shown to provide observable evidence of learning in collaborative learning settings \cite{roschelle1995construction}.}

\subsection{Cybersecurity Education in the AIED Context}

Cybersecurity is a key component of contemporary computing education, integrating technical systems with human, informational, and organizational concepts to protect operations against adversarial threats~\cite{cc2023,jtf-csec2017}. Alongside rapidly expanding areas such as artificial intelligence, cybersecurity has gained increasing prominence, and mastery of cybersecurity competencies is now widely viewed as core preparation for graduates in computer science and related fields~\cite{cc2023,Raj2022competencies}.

Despite its importance, cybersecurity remains marginal in AIED and learning analytics research. A systematic review found only 35 relevant studies among more than 3,000 publications, most using primarily shallow descriptive measures~\cite{Svabensky2022applications}. Furthermore, most past papers on the topic focused on privacy or infrastructure rather than cybersecurity learning itself~\cite{Asatryan2024,Drachsler2016}.

Existing research on cybersecurity education within AIED and learning analytics focused on privacy awareness, informal learning environments, program-level analyses, and online courses~\cite{Franco2023,Brennan2022,Kitto2020,Asif2015,Vogelsang2015}. None of these studies examined students in dedicated higher-education cybersecurity degree programs. Although cybersecurity education has gained visibility in computing education research~\cite{Svabensky2020what}, it remains rare within AIED. This study addresses that gap by presenting learning analytics evidence from a higher education program centered on cybersecurity.

\subsection{Tabletop Exercises (TTXs)}
\label{subsec:related-work-TTX}

Tabletop exercises (TTXs) are a form of experiential learning in which small groups collaboratively work through complex, time-constrained scenarios in a shared instructional setting~\cite{Vykopal2024research,Angafor2024}. Instructors issue common prompts, while teams develop responses independently and submit outcomes orally or through digital systems of varying sophistication. This structure supports coordinated engagement without requiring fully synchronized group activity.

TTXs help learners practice responses to realistic, high-stakes situations that are difficult to reproduce in classrooms, including disaster response, public health emergencies, healthcare crises, and cybersecurity, which is the focus of the present study~\cite{Kavrestad2025}. More broadly, TTXs belong to the tradition of simulation-based pedagogy, which is well established in AIED and learning analytics research~\cite{Yan2023simulation}. 
\revisiontwo{Compared to technical exercises in cyber ranges -- another popular approach for realistic cybersecurity simulations~\cite{Vykopal2021scalable} -- TTXs are much more lightweight and relevant also for less technically-oriented roles (e.g., IT management).}

\section{The INJECT Exercise Platform}

Tabletop exercises were traditionally conducted offline using printed materials, which imposed substantial preparation demands on instructors and required labor-intensive manual assessment. In response, digital tools have emerged to support TTX-based instruction, as documented in a recent survey~\cite{Vykopal2024research}. Our study uses the open-source INJECT Exercise Platform (IXP)~\cite{Svabensky2024from}, a browser-based system supporting the full exercise lifecycle while reducing instructor workload. IXP provides scenario-critical information, is freely available and actively maintained, and has been refined over time through sustained feedback from instructors.

\subsection{Educational Goals and Learning Objectives}
\label{subsec:inject-ttx-goals}

While TTXs apply across many instructional domains (Section \ref{subsec:related-work-TTX}), we focus on cybersecurity. Here, TTXs enable learners to practice responses to cyber crises that disrupt IT operations, such as data breaches. Teams collaboratively address technical incidents alongside communication and coordination challenges, developing competencies in conditions that mirror professional practice and support workforce readiness in a rapidly evolving field~\cite{cc2023,Raj2022competencies}.

The exercises model the work of a Computer Security Incident Response Team (CSIRT) in medium- to large-scale organizations. Participants engage in open-ended scenarios that require prioritizing and resolving multiple, simultaneous incidents under time constraints of roughly 90 minutes, reflecting the pace and uncertainty of real incident response. Learning outcomes align with the Incident Response role in the NICE Cybersecurity Workforce Framework~\cite{NICE}. In addition to technical skills, the activities emphasize professional dispositions~\cite{cc2023,Raj2022competencies}, including teamwork, decision-making under pressure, and effective communication within the CSIRT and with external stakeholders.
\revisiontwo{Specific tasks that the teams address include investigating connections from suspicious IP addresses to an organization's infrastructure, advising non-expert users affected by cyber attacks, and writing up a report documenting the findings.}

\subsection{How Students Learn Using the INJECT Exercise Platform}

\begin{figure}[!ht]
    \centering
    \includegraphics[width=\textwidth]{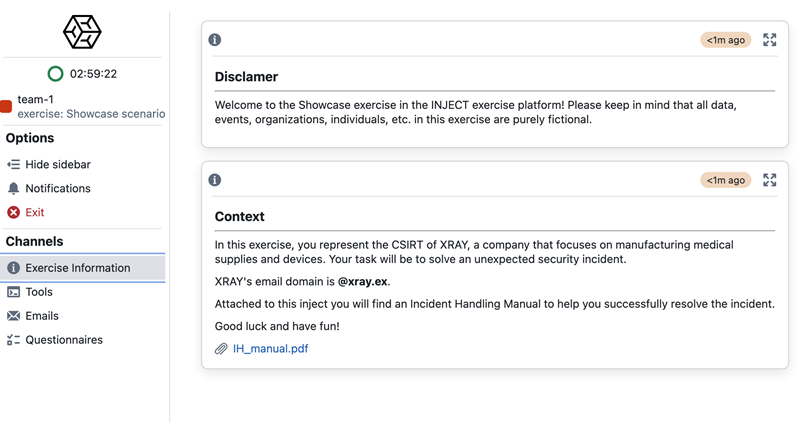} 
    \caption{IXP from the student perspective, displaying recent injects.}
    \label{fig:IXP-screenshot}
\end{figure}

Tabletop exercises unfold through \textit{injects}, predefined messages released during the activity to introduce new information and advance the scenario~\cite{Svabensky2024from}. An inject may, for example, inform students of a breach in their simulated organization. In traditional TTXs, instructors typically deliver such messages verbally or on paper. The INJECT Exercise Platform instead automates inject delivery, presenting them directly within each team's web interface, as shown in Figure~\ref{fig:IXP-screenshot}.

Upon receiving an inject, teams decide how to respond and enact their decisions through the platform (Figure \ref{fig:IXP-screenshot}). These tools abstract real-world applications, allowing teams to carry out actions that resemble professional practice and receive immediate feedback. Professional communication is also central to TTXs, and in IXP, this is modeled through emails from simulated stakeholders that pose questions or requests requiring deliberation. Teams must compose appropriate replies under time pressure as the incident unfolds. \revision{Responses to these emails require both lower- and higher-order cognitive processes, such as confirming information or applying procedures (Table \ref{tab:codebook}).}
\revisiontwo{Email is the only tool for communicating with simulated in-exercise actors. Spoken communication between team members is also relevant but outside the scope of our analysis.}

Every team action, including tool use and email communication, is logged with full metadata and microsecond-resolution timestamps in JSONL format. These logs support the identification of milestones that mark key events in the TTX and allow instructors to monitor progress and, if desired, assign grades.

\section{Methods}

\subsection{Participant Sample}

\revisiontwo{By including participants from multiple nations and contexts, we aim to improve the external validity of our findings~\cite{handbook-CER1}.} In \revisiontwo{the Czech Republic}, four sequential tabletop exercises (TTX0–3) were run as a capstone in a semester-long cybersecurity incident response course at a public university, involving 36 computing majors per exercise, organized into 13 skill-balanced teams (10 teams of three and three teams of two). The exercises were spaced out across several weeks. In \revisiontwo{Estonia}, a fifth exercise (TTX4) included 11 teams at a cross-border cybersecurity event: five university teams, five vocational-school teams, and one educator team.

In late 2024, \numteamstotal\ teams with \numstudentstotal\ learners participated. One team declined research consent, leaving \numteamsanalyzed\ teams and \numstudentsanalyzed\ learners for analysis. The sample is comparable to or larger than prior studies of collaborative cybersecurity work~\cite{Won2024Cybersecurity}. \revision{All teams in our sample completed up to 18 milestones in a fixed order (though multiple milestones can be worked on simultaneously), limiting task difficulty confounds. Finally, 68 (20\%) milestone completions included no email.} \revisiontwo{Such milestones mostly included simple tasks such as blocking an IP address, which did not require communication with the simulated exercise stakeholders.}

\subsection{Coding of Emails}

Annotating emails is substantially more complex than annotating exercise objectives, which are few and brief, and therefore requires dedicated qualitative coding. We developed a codebook to classify security and incident-response emails by the highest cognitive demand imposed on the recipient. \revision{We ensured that the scheme was sufficiently grounded in Bloom's taxonomy by initially starting codebook development based on ACM CCECC's Bloom's taxonomy for computing~\cite{10.1145/3587276}}. The codebook defines six categories spanning simple acknowledgments to higher-order demands involving analysis, evaluation, or creation; Table~\ref{tab:codebook} presents the categories, definitions, and representative examples.
 
\begin{table}[htpb]
\centering
\caption{Overview of the Codebook categories.}
\label{tab:codebook}
\setlength{\tabcolsep}{2mm}
\resizebox{\textwidth}{!}{
\begin{tabular}{lp{6.1cm}p{4cm}}
\toprule
\textbf{Category} & \textbf{Definition} & \textbf{Example} \\
\midrule
Remembering & Pure acknowledgement or status updates with no action required. & ``Thanks for the info.'' \\
\addlinespace
Understanding & Provide trivial self-facts or confirm information already known. & ``What is your email?'' \\
\addlinespace
Applying & Execute known steps or fetch specified resources. & ``Send 48h access logs for \texttt{nice-project.uni.ex}.'' \\
\addlinespace
Analyzing & Investigate without prescribed steps; decide what data or methods to use. & ``What is going on? Please provide more information.'' \\
\addlinespace
Evaluating & Make judgments under uncertainty (policy, impact, authenticity). & ``Should we inform NCISA about this incident?'' \\
\addlinespace
Creating & Produce new artifacts such as summaries, advisories, or drafts. & ``Please draft the public statement about the breach.'' \\
\bottomrule
\end{tabular}
}
\end{table}

Three independent coders applied the codebook to individual emails from the INJECT Exercise Platform using a three-stage process of independent coding, discussion-based reconciliation, and final consolidation. \revision{When text reflected multiple Bloom levels, coders assigned the highest level; when no emails were present, no code was assigned.} Inter-rater reliability was moderate to substantial. Pairwise Cohen's $\kappa$ ranged from 0.555 to 0.678 across coder pairs, and Fleiss' kappa for three raters on complete cases was 0.636. Reliability varied by category, with higher agreement for Applying and Creating, moderate agreement for Analyzing, Remembering, and Evaluating, and the lowest agreement for Understanding. Given the inherent ambiguity of cognitive constructs, $\kappa$ values around 0.6 are commonly considered acceptable in related research \cite{levin2022evaluating,karpen2016assessing}. We provide anonymized analysis and preprocessing code via a public Git repository.\footnote{\url{https://github.com/conradborchers/bloom-ttx/}} The study data are also open source \cite{ttx_dataset}.

\section{Study 1: Explaining Performance via Bloom (RQ1)}

Study 1 tested whether (a) Bloom-coded cognitive demand in team emails predicted milestone success and whether (b) misalignment between required and observed Bloom levels predicted outcomes.

\subsection{Method}

For each team and milestone, we coded the highest Bloom level evident in the related email exchange. Milestone completion was a binary outcome ($1$ for completion, $0$ for non-completion). We computed a discrepancy score as the distance between the Bloom level required and that observed, coded as $0$ for exact matches, $1$ for adjacent levels, and $2$ otherwise. \revision{Missing data were coded as 2, since no response is conceptually similar to an inadequate response.} 

We fit a sequence of generalized linear mixed models predicting the log-odds of milestone achievement. All models included a random intercept for each team to account for repeated observations. The baseline model included the observed Bloom level as a fixed effect. The second model added the discrepancy score to capture misalignment between required and enacted cognitive levels. The final model further included the interaction between Bloom level and discrepancy to test whether the effect of misalignment depended on the absolute cognitive level at which teams were communicating.

Models were compared sequentially using likelihood-ratio tests, retaining the best-fitting specification. Fixed effects are reported as odds ratios with 95\% confidence intervals, which quantify multiplicative changes in the odds of success in logistic models. An odds ratio above one indicates increased odds, while values below one indicate reduced odds, for a one-unit increase or category presence. Odds ratios describe relative changes rather than absolute probabilities.

\subsection{Results}

Adding discrepancy significantly improved fit over a Bloom-only baseline, $\chi^2(1)=7.18$, $p=.007$, whereas the Bloom-discrepancy interaction did not, $\chi^2(2)=0.45$, $p=.800$. In the selected additive model, greater discrepancy reduced the odds of success (OR = 0.65, 95\% CI [0.48, 0.90], $p=.008$), and Bloom category was no longer significant ($p$'s $>.077$). Between-team variability was substantial (ICC = .20), but fixed effects explained little variance (marginal $R^2=.04$; conditional $R^2=.23$), indicating limited explanatory power of Bloom alignment.

A complementary $\chi^2$ test across discrepancy levels (0–2) showed a significant but weak association with milestone achievement, $\chi^2(2, N=340)=6.41$, $p=.041$, Cramér's $V=.14$, suggesting that smaller discrepancies were associated with higher success rates but with modest practical impact.

\section{Study 2: What Predicts Performance? (RQ2)}

The second study examined which feature families best predict milestone achievement. Moving beyond Study~1's theoretical focus, we compared models using Bloom-coded email categories, linguistic features from team communication, and log-based behavioral indicators, alone and in combination, to assess whether Bloom coding adds predictive value beyond automated text and log traces.

\subsection{Method}
\label{subsec:method-study2}

In the second study, we fused system logs and email with Bloom-coded indicators into text sequences and numeric features. All timestamps were converted to UTC and aligned by \texttt{teamID}. For each milestone, we aggregated logs and emails since the previous milestone or session start. Emails were concatenated with a delimiter token (\texttt{<|CHAIN|>}). We also computed counts of logs and emails, action-type frequencies, average log and email lengths, elapsed time since the prior milestone, email response times, number of unique senders, and Bloom code frequencies. Together, these features constituted the milestone-level indicators.
We grouped the resulting predictors into three groups:
\begin{itemize}
    \item \textbf{Text features:} embeddings of \texttt{chained\_logs} and \texttt{chained\_emails} generated using the \texttt{all-MiniLM-L6-v2} model from \texttt{SentenceTransformers}.
    \item \textbf{Log features:} numeric aggregates derived from system interactions, excluding Bloom-coded variables.
    \item \textbf{Bloom features:} counts of Bloom taxonomy codes (e.g., \textit{remembering}, \textit{applying}) extracted from coded emails within objectives.
\end{itemize}

\paragraph{Evaluation design.}
To assess predictive value, we trained $\ell_{1}$-regularized logistic regression models using different feature combinations, with milestone achievement as the binary outcome. Data were split into 80\% training and 20\% test sets, stratified by outcome and grouped by \texttt{teamID}, with 5-fold \texttt{GroupKFold} cross-validation on the training set. Inputs were standardized. $\ell_{1}$ regularization mitigates overfitting and enforces feature selection \revision{(by shrinking irrelevant coefficients to 0 and effectively dropping them out of the model)}. Performance was evaluated using AUC with 95\% confidence intervals, computed across folds and via stratified bootstrap on the test set.

\paragraph{Interpretability.}
Interpretation relied on coefficients from the $\ell_{1}$-regularized logistic regression. With standardized features, the coefficient sign and magnitude directly reflect the direction of the association with milestone achievement. Accordingly, post-hoc explainers \revision{designed for complex, non-linear architectures} (e.g., SHAP/LIME) were unnecessary, reducing overfitting risk and avoiding known inconsistencies in these interpretability methods \cite{swamy2022evaluating}. We report the ten largest absolute weights, spanning text embeddings and engineered indicators.

\subsection{Results}

Table~\ref{tab:study2-results} reports predictive performance across feature sets. Bloom-coded features alone performed at near-chance levels, whereas text and log features were substantially more predictive. Combining text and log features yielded the strongest results (test AUC = 0.80), indicating a complementary signal, with Bloom features adding little incremental value beyond embeddings.

\begin{table}[htpb]
\centering
\caption{Cross-validated and test set performance for different feature sets.}
\label{tab:study2-results}
\setlength{\tabcolsep}{1.5mm}
\begin{tabular}{lcccc}
\toprule
\textbf{Model} & \textbf{CV AUC} & \textbf{CV 95\% CI} & \textbf{Test AUC} & \textbf{Test 95\% CI} \\
\midrule
Bloom only            & 0.551 & [0.468, 0.634] & 0.541 & [0.469, 0.615] \\
Text only             & 0.740 & [0.720, 0.760] & 0.727 & [0.666, 0.786] \\
Log only              & 0.714 & [0.667, 0.761] & 0.727 & [0.669, 0.783] \\
Text + Bloom          & 0.740 & [0.720, 0.760] & 0.727 & [0.666, 0.786] \\
Log + Bloom           & 0.714 & [0.667, 0.761] & 0.727 & [0.669, 0.783] \\
Text + Log            & 0.790 & [0.760, 0.821] & 0.804 & [0.747, 0.856] \\
All (Text + Log + Bloom)  & 0.790 & [0.760, 0.821] & 0.804 & [0.747, 0.856] \\
\bottomrule
\end{tabular}
\end{table}

\begin{figure}[htpb]
    \centering
    \includegraphics[width=.91\linewidth]{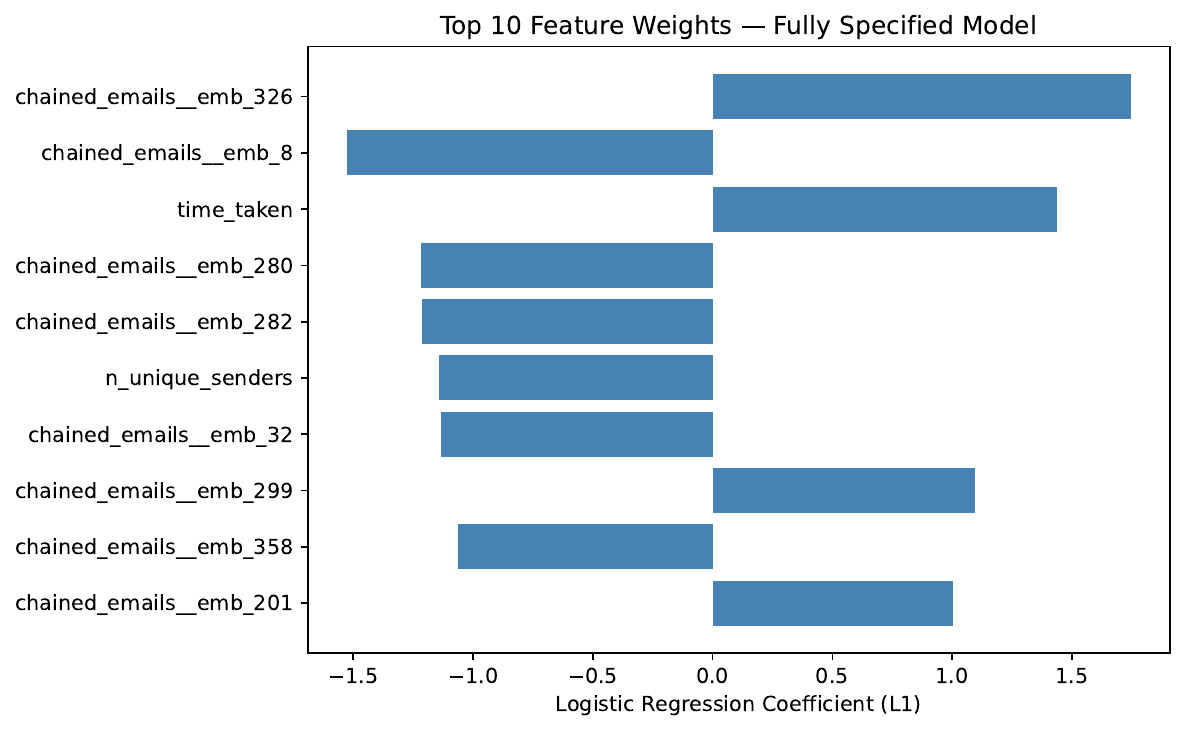}
    \caption{Top ten features in the fully specified model (Text+Log+Bloom). Positive weights indicate higher predicted probability of milestone achievement. \revision{Subscripts for embedding-based features represent the n-th dimension of the vector.}}
    \label{fig:feature-weights-top10}
\end{figure}

Figure~\ref{fig:feature-weights-top10} displays the ten strongest predictors from the full model. Most derive from \texttt{chained\_emails} embeddings, reinforcing the predictive value of email communication, with additional contributions from temporal and structural features. The coefficient sign indicates the direction of association with milestone achievement.

\section{General Discussion}

Instructional alignment has long been treated as a central principle of learning design~\cite{biggs1996enhancing,porter2002measuring}, yet it is rarely operationalized in authentic, data-rich contexts such as SBL. Most prior work examines alignment through curriculum artifacts or rubric-based analyses~\cite{Ayyanathan_2022}, or reviews scaffolding in simulations without measuring alignment as it occurs during activity~\cite{Chernikova2025}.

This article addresses this gap by integrating theory-driven coding with predictive modeling of multimodal team data. Study~1 frames alignment as a discrepancy between instructional objectives and enacted communication, while Study~2 compares the predictive value of Bloom-coded features with text and log data. Together, the studies connect interpretable, theory-based measures of alignment with scalable analytics that support accurate prediction.

\subsection{Study 1: Alignment and Milestone Success}

From a constructive alignment perspective, Study 1 results explain why discrepancy, rather than nominal Bloom levels, predicts success. Biggs' account of alignment emphasizes that learning improves when activities and assessment elicit the same cognitive work as the intended objectives \cite{biggs1996enhancing}. Our finding that misaligned teams underperform extends this principle to simulation-based contexts using trace data. Prior meta-analyses of scenario-based learning similarly show that scaffolds improve outcomes by aligning what learners actually do with the targeted cognitive demands, not by merely increasing task difficulty \cite{Chernikova2025,chernikova2020simulation}. In tabletop exercises, where reasoning is enacted primarily through time-pressured communication~\cite{Vykopal2024research,Svabensky2024from}, a Bloom gap likely reflects coordination without the intended analytic, evaluative, or generative thinking, consistent with work distinguishing surface participation from deeper cognition \cite{wine2023reinforcing}.

For TTX design, these results suggest prioritizing scaffolds that help teams maintain reasoning at the intended cognitive level rather than adding more complex injects. Because TTXs generate rich communication traces, they enable real-time alignment diagnosis by combining transcription, automated classification, and predictive modeling. Comparable methods have captured self-regulated learning processes as they unfold \cite{borchers2025large}. Applied to TTXs, such approaches could enable timely prompts or feedback when misalignment emerges, turning Bloom discrepancy from a post-hoc indicator into a mechanism for just-in-time support in high-pressure learning environments.

\subsection{Study 2: Predictive Value of Feature Families}

Study~2 frames Bloom codes as interpretable diagnostics while showing that linguistic and interaction traces are more predictive of performance. This mirrors prior discourse-analytic and learning analytics work demonstrating that models grounded in communication structure, language use, and temporal participation outperform coarse categorical labels, and that combining multiple trace types yields complementary value \cite{suraworachet2024predicting,borchers2025combining,wong2025rethinking}. Our results hence support combining fine-grained behavioral and linguistic signals for forecasting performance.

The limited contribution of Bloom frequencies beyond embeddings \revision{could indicate limited value for our prediction task. Alternatively, it could indicate} that their information is already captured in the semantic structure of communication traces. This aligns with evidence that large language models achieve strong classification performance while reducing interpretability \cite{vajjala2025text}. Practically, this suggests a division of labor: Bloom coding supports explanation and design, whereas embeddings and engineered features support real-time risk detection. Similar complementarities between rich interaction logs and textual analysis have been reported in large-scale cyber and simulation-based exercises \cite{pfaller2025data,suraworachet2024predicting}.

Across both studies, Bloom discrepancy helps explain why teams succeed or struggle, while multimodal features indicate who is likely to do so and when. For AIED, this suggests a pipeline in which predictive models flag at-risk moments using text and log traces, and alignment analytics provide cognitively meaningful rationales for feedback and redesign. \revision{Analytics related to this feedback could be directly shown to students to enhance learning or to instructors to improve instruction accompanying the simulation-based training or curricular redesign.} This synthesis responds to calls for analytics that are simultaneously accurate and pedagogically actionable \cite{wise2014designing}, and it links platform-level telemetry with discourse-level evidence of thinking in TTXs and related simulations. Future work should extend these diagnostics beyond Bloom categories toward network-based representations, such as epistemic network analytics, to support more nuanced and instructionally useful insights \cite{borchers2024revealing}.

\subsection{Limitations and Future Work}

Despite diverse data spanning two countries, our findings are bounded by the specific cybersecurity platform and TTXs, each lasting approximately 1.5 hours. Broader validity will require replication across domains and longer SBL formats, ideally incorporating milestone-level difficulty controls and alternative operationalizations that model reasoning trajectories rather than peak Bloom levels. \revision{Our results are also correlational in nature, calling for quasi-experimental studies potentially matching groups with different Bloom discrepancies by prior knowledge, ability, or similarly adjusting for milestone difficulty.} 

We prioritized interpretable linear models, but more complex architectures could improve predictive performance\revision{, and so could more complex measures of Bloom discrepancy in larger samples, such as penalizing larger discrepancies more strongly instead of treating all Bloom taxonomy gaps as equally-spaced}. Future work should also examine hybrid approaches that pair high-performing multimodal models with explanations instructors can readily use. \revision{We also acknowledge that more qualitative analysis will be needed to make embedding-based features consumable for students and teachers \revisiontwo{(e.g., by sampling prototypical examples corresponding to changes in salient embedding dimensions \cite{borchers2025disentangling})}.} \revisiontwo{Finally, correlating milestone performance with assessments of learning (e.g., via external pre-post tests) as well as studying fine-grained process measures of remote collaboration (e.g., email revisions) are subject to future work.}

\section{Conclusion}

We operationalize constructive alignment as a trace-based measure for simulation learning and show how it complements multimodal prediction by integrating linguistic communication (email/text embeddings) and behavioral interaction logs. In cybersecurity tabletop exercises, alignment, measured as Bloom-level discrepancy between intended objectives and enacted communication, predicted milestone success, whereas nominal Bloom level did not. At the same time, linguistic and interaction traces outperformed Bloom frequencies in prediction, separating explanation from detection: alignment explains why teams succeed or fail, while multimodal features indicate who is at risk and when.

This supports an actionable AI-based pipeline: text and log models flag impending milestone failures, and alignment diagnostics guide targeted, cognitively grounded support such as prompts to analyze or justify. Although evaluated in cybersecurity TTXs with human-coded labels, the approach generalizes to other domains, semi-automated coding, and embedded interventions triggered by misalignment. By combining predictive accuracy with explanatory clarity, this work advances AIED toward timely, pedagogically meaningful guidance.

\begin{credits}
\subsubsection{\ackname}
This research was supported by the Open Calls for Security Research 2023--2029 (OPSEC) program granted by the Ministry of the Interior of the Czech Republic under No. VK01030007 -- Intelligent Tools for Planning, Conducting, and Evaluating Tabletop Exercises.
\end{credits}

\bibliographystyle{splncs04}
\bibliography{main} 

\end{document}